\documentstyle[preprint,aps,psfig]{revtex}
\begin{document}
\preprint{SNUTP 98-025}
\title{ Thermodynamic Curvature of the BTZ Black Hole}
\author{Rong-Gen Cai\thanks{email: cai@atlantis.snu.ac.kr} and
Jin-Ho Cho\thanks{email: jhcho@galatica.snu.ac.kr}}
\address{Center for Theoretical Physics, Seoul National University,
Seoul 151-742, Korea}

\maketitle

\begin{abstract}
Some thermodynamic properties of the 
Ba\~nados-Teitelboim-Zanelli (BTZ) black hole are studied to get the 
effective dimension of its corresponding statistical model.
For this purpose, we make use of the geometrical approach to the 
thermodynamics: Considering the black hole as a thermodynamic system 
with two thermodynamic variables (the mass $M$ and the angular 
momemtum $J$), we obtain two-dimensional Riemannian thermodynamic
geometry described by positive definite Ruppeiner metric. 
From the thermodynamic curvature we find that the extremal 
limit is the critical point. The effective spatial dimension 
of the statistical system corresponding to the near-extremal BTZ 
black holes is one. Far from the extremal point,
the effective dimension becomes less than one, which leads to one
possible speculation on the underlying structure for the 
corresponding statistical model.
\end{abstract}

\section{Introduction}
Since the works of Bekenstein \cite{Bek}, Bardeen, 
Carter and Hawking \cite{Bar}, and Hawking \cite{Haw}, black holes 
have been known to have the ordinary thermodynamical properties.  
There are four laws of black hole mechanics analogous to the four
laws of ordinary thermodynamics. (The surface gravity on the event 
horizon can be interpreted as the temperature of the black hole. 
One quarter of the event horizon area corresponds to the entropy of 
the black hole.) These correspondences proved not just to be an 
analogy but to be a realization of thermodynamics involving 
thermal radiation \cite{Haw}. 
Now, it is widely believed that a black hole is a thermodynamic 
system.  

The statistical interpretation of the black hole entropy has been
one of the most fascinating subjects to understand the microstates
behind its thermodynamics. So far, many progresses have been made
for some specific models (extremal and near extremal black holes) 
in the string theory with the help of supersymmetry 
\cite{Vaf,Hor}.
Another success was made for the BTZ black hole \cite{BTZ}
making use of conformal symmetry on the boundary 
\cite{Car1,Stro}. With the notion 
that many supergravity solutions of superstring theory can be 
related to the lower dimensional anti-de Sitter gravity solutions
via U-duality \cite{Hyun}, one can think these two approches might 
not be quite 
remote from each other. In fact recently, duality between string 
theory on various anti-de Sitter spacetimes and various conformal 
field theories was conjectured \cite{Mal,Pol,Wit}.
In this sense it becomes very important to find the conformal field 
theory corresponding to each black hole to account for its statistical 
origin of the entropy and the BTZ black hole will be the corner stone
along this line.

Recently, it was reported for some near extremal cases including BTZ 
black hole, the relevant degrees of freedom concerned with the entropy 
live effectively on one spatial dimension \cite{Low,Gho}. 
This means that their 
corresponding conformal field theory must be $(1+1)$-dimensional.
(So far, there have been many controversies on the exact place 
where the degrees of freedom reside, i.e., whether it is the interior
of the black hole or else the horizon. According to the recent
results above, it is likely that the exact place is concerned with 
just one spatial 
dimension and that BTZ geometry partially constitutes the internal
space for a class of black holes).
In this paper, we ellaborate on this aspect for the BTZ black hole.
For this purpose, we make use of the geometrical approch to the 
thermodynamical system, which is usually adopted in the ordinary 
thermodynamic fluctuation theory \cite{Wei,Rup1}. 
In the case at hand, the 
thermodynamic geometry is described in terms of two dimensional 
Ruppeiner metric (see \cite{Rup} for the details on the thermodynamic
geometry). 
We find that the thermodynamic curvature of the BTZ black holes is
always positive and diverges strongly as the black hole approaches 
its  extremal limit. This divergence is in agreement with the 
existence of the critical points, on which many authors have 
suggested in different contexts \cite{Cur,Pav,Tra,Kar,Cai2,Cai3}.
We find from the thermodynamic curvature that the effective 
spatial dimension of the statistical system corresponding to the BTZ 
black hole is `one' in the extremal limit. However, this result
does not continue to the region far from the limit, for which 
the effective dimension becomes less than one. This strange result 
suggests that the corresponding statistical model is composed
of some extended objects rather than of point particle gas.

This paper is organized as follows: In section II, we give a 
brief review on the geometrical method on the thermodynamics.
In section III, we apply this approach to the BTZ black hole and 
get the Ruppeiner metric and observe that the extremal black hole
corresponds to the critical point.
In section IV, the thermodynamic
curvature and the effective spatial dimension of the corresponding
statistical system is calculated. 
Section V concludes the paper with the discussion on the results. 

\section{Geometrical Method for Thermodynamics}
It was Weinhold \cite{Wei} who first introduced the geometrical concept 
into the thermodynamics. He considered a sort of Riemannian metric,
i.e., positive definite metric in 
the space of the thermodynamic parameters.
The Weinhold metric $W_{\mu\nu}$ is defined as
the second derivatives of the internal energy $U$ with respect to the 
entropy and other extensive variables of a thermodynamic system 
$N^\mu\equiv(S,N^i)$, 
\begin{equation}
\label{weinmetric}
(d N^\mu,\,d N^\nu)\equiv W_{\mu\nu}=
\frac{\partial^2 U}{\partial N^\mu\partial N^\nu}.
\end{equation}
In his picture, the second law of thermodynamics can be derived
just from the Schwarz inequality \cite{Wei}. 
\begin{equation}
(d N^\alpha,\,d N^\alpha)(d N^\beta,\,d N^\beta)
-(d N^\alpha,\, d N^\beta)^2\geq 0.
\end{equation}
For example, the above inequality for the two vectors,
the entropy $S$ and its conjugate, 
the temperature, $T$, result in the stability condition $C_p\geq C_v$,
where $C_p$ and $C_v$ are heat capacities at constant pressure and 
constant volume respectively.
However, the geometry based on this metric was considered 
to have no physical meanings in the context of purely 
equilibrium thermodynamics \cite{Rup}.

In 1979, Ruppeiner \cite{Rup1} introduced a metric, which is  
defined as the second derivatives of the entropy with respect to the 
internal energy and other extensive variables of a thermodynamic
system, $a^\mu\equiv(U,a^i)$, 
\begin{equation}
\label{rupmetric}
S_{\mu\nu}=-\frac{\partial^2 S}{\partial a^\mu\partial a^\nu}.
\end{equation}
Although the Weinhold metric and 
Ruppeiner metric are conformally related with the temperature $T$
as the conformal factor, 
\begin{equation}
W_{\mu\nu}dN^\mu dN^\nu=TS_{\mu\nu}da^\mu da^\nu,
\end{equation}
the Ruppeiner geometry has its physical meaning 
in the equilibrium thermodynamic fluctuation
theory. The components of the inverse Ruppeiner metric give us  
second moments of fluctuations.
The advantage of this geometrical concept is that the fluctuation 
theory can be described in the covariant way so that for any set of
thermodynamical variables, all the results can be obtained in the 
same way.

Since the proposal of  Ruppeiner, many investigations on 
the geometric meanings of the Ruppeiner metric have been carried 
out in various thermodynamic systems, such as the 
ideal classical gas, ideal paramagnet,
multicomponent ideal gas, ideal quantum gases, 
Takahashi gas, one-dimensional Ising model, van der Waals model 
and so on. In particular, the Riemannian scalar curvature of the
Ruppeiner metrics has been calculated for many thermodynamic
systems. It turns out that
the Riemannian scalar curvature $\Re$ of the Ruppeiner geometry is
related to the correlation volume as,
\begin{equation}
\Re=\kappa_2\xi^{\bar{d}}, \label{scale}
\end{equation}
where $\kappa_2$ is a constant with absolute value of order unity,
$\xi$ is the correlation length, and $\bar{d}$ is the spatial 
dimension of the statistical system (not to be confused with the 
dimensionality of the Riemann geometry, which is the 
the number of thermodynamic quantities). 
For example, the Riemannian curvature is zero for single-component
ideal gas, which implies that there is no 
interaction (including the statistical interaction) in the system. 
For ideal quantum Fermion gas,
the thermodynamic curvature is always positive in the sign
convention of \cite{Rup} which we will adopt, while for the ideal 
quantum Boson gas, 
it is always negative and diverges strongly as the temperature 
approaches zero. This phenomenon is related to the Bose-Einstein 
condensation. Near the critical point, the Riemannian curvature 
always has an infinite divergence (for a detailed review see 
\cite{Rup}) These properties are in accord with those of the
correlation length. One important thing to note is that the
relation (\ref{scale}) is valid even far from the critical point, 
although seemingly it looks like some kind of scaling behavior. 
Another important
physical meaning of the scalar curvature is that it provides the 
lower bound on the volume for which the classical fluctuation
theory is valid. If we take some sample volume of smaller size than 
this bound, we cannot apply the classical fluctuation theory.

\section{Ruppeiner Metric of BTZ Black Hole}
\subsection{BTZ Black Hole}
The BTZ black hole is the solution of the (2+1)-dimensional 
Einstein gravity with a negative cosmological constant. 
The action describing the black hole is \cite{BTZ} 
\begin{equation}
I=\frac{1}{2\pi}\int d^3x \sqrt{-g} (R+2l ^{-2}),
\end{equation}
where $R$ denotes the scalar curvature (we hope this not to be 
confused with the thermodynamic curvature $\Re$ discussed in the 
previous 
section) and $l^{-2}$ represents the negative cosmological constant. 
The BTZ black hole metric is
\begin{equation}
\label{btz}
ds^2=-N(r)dt^2+N^{-1}(r)dr^2+r^2(N^{\phi}(r)dt+d\phi)^2,
\end{equation}
where
\begin{equation}
N(r)=-M+\frac{r^2}{l^2}+\frac{J^2}{4r^2},\ \
 {\rm and}\ \ N^{\phi}(r)=-\frac{J}{2r^2},
\end{equation}
$M$ and $J$ are two integration constants, which can be interpreted as 
the mass and angular momentum of the black holes, respectively. 
Throughout this paper, the units $8G=c=\hbar =k_B=1$ will be used.

BTZ black hole (\ref{btz}) can be constructed by orbifolding
the (2+1)-dimensional anti-de Sitter space with the spacelike 
Killing vector fields. 
Therefore the BTZ black 
hole is locally a constant curvature spacetime. Through the Einstein 
field equations $R_{\mu\nu} -\frac{1}{2}Rg_{\mu\nu}=l^{-2}g_{\mu\nu}$,
the scalar curvature of the BTZ black hole spacetime is 
 \begin{equation}
 R=-6l^{-2}.
 \end{equation}
The BTZ black hole has two horizons
\begin{equation}
r^2_{\pm}=\frac{1}{2}Ml^2( 1\pm \triangle), \ \
\triangle=[1-(J/Ml)^2]^{1/2}.
\end{equation}
It is obvious that $J \le Ml$ and $M\ge 0$ should be satisfied
in order that the metric (\ref{btz}) to have the black hole structure. 
The Hawking temperature  $T$ of the hole is readily obtained 
\begin{equation}
T=\frac{r^2_+-r^2_-}{2 \pi r_+l^2}=\frac{M\triangle}{2 \pi r_+}.
\end{equation}
The Bekenstein-Hawking entropy is
\begin{equation}
S=4\pi r_+=4\pi \left[\frac{M}{2}l^2(1+\triangle)\right ]^{1/2}.
\end{equation}
These thermodynamic quantities obey the first law of thermodynamics
\begin{equation}
\label{firstlaw}
dM=TdS+\Omega _H dJ,
\end{equation}
where $\Omega _H=J/2r_+^2$ is the angular velocity of the hole. 
 Another feature of BTZ black holes is that its heat capacity is 
always positive, which is in contrast with the Schwarzschild case.  
According to the formula 
$C_J=(\partial M/ \partial T)_J$, we have \cite{Cai3}
\begin{equation}
\label{capacity}
C_J=\frac{4  \pi  r_+ \triangle}{2-\triangle}.
\end{equation}
Because of $0\le \triangle \le 1$,  
$C_J\ge 0$ always holds, which 
 means the temperature to increase with the mass. Therefore, 
 the BTZ black hole can be in stably thermal
equilibrium with an arbitrary volume heat bath. When $\triangle =1$, 
i.e.  $J=0$, we have $C_J=4 \pi l \sqrt{M}$. And when $\triangle =0$,
i.e. $J=M l$, we have $C_J=0$, corresponding to the extremal BTZ black 
holes. In that case, the two horizons of the hole coincide, and the 
Hawking temperature becomes zero.

Since the discovery of the BTZ black holes, many works 
have been done on the classical and quantum properties 
of BTZ black holes. 
The statistical explanation of entropy was first provided by 
Carlip \cite{Car1} and the basic points are spelled out in
the recent paper \cite{Stro}. 
A nice review about the physics of BTZ black holes can be 
found in \cite{Car}.

\subsection{Thermodynamic Metric}
Now we construct the thermodynamic geometry of BTZ black hole based 
on its thermodynamics. 
According 
to the definition (\ref{weinmetric}), the Weinhold metric of BTZ
black holes can be written down as
\begin{eqnarray}
\label{wei}
ds^2_{\rm W} & \equiv & \left  ( \frac{\partial ^2 M}{\partial S^2}
           \right )_JdS^2   + \left(\frac{\partial ^2M}{\partial J^2}
         \right)_SdJ^2 \nonumber \\
      &=&\frac{T}{C_J}dS^2 +I_S dJ^2,
\end{eqnarray}
where mass $M$ corresponds to the internal energy $U$ while entropy
$S$ and angular momentum $J$ are taken as the extensive variables 
$N^\mu$. $C_J$ is given by
Eq.~(\ref{capacity})  and $I_S$ is defined as
\begin{equation}
I_S  \equiv \frac{1}{\Omega _H} 
\left(\frac{\partial \Omega _H}{\partial
 M}\right )_S=\frac{1}{M} +\frac{Ml^2\triangle (1+\triangle)}{J^2}.
\end{equation}
It is easy to see that the Weinhold metric is regular even for the 
extremal BTZ black holes. Our interest is the Ruppeiner metric.
Rewriting Eq.~(\ref{firstlaw}), we have
\begin{equation}
dS=\beta dM -\mu dJ,
\end{equation}
where $\beta =1/T$ and $\mu=\beta \Omega _H$. According to the
definition (\ref{rupmetric}), the Ruppeiner metric of the BTZ black 
holes in the $a=(M, J)$ coordinates is
\begin{eqnarray}
\label{rm1}
ds^2_{\rm R} &\equiv & -\left(\frac{\partial ^2 S}{\partial M^2} 
      \right)_J  dM^2
     -\left(\frac{\partial ^2 S}{\partial J^2}\right)_M 
dJ^2 \nonumber\\
     &=&\frac{1}{T^2 C_J}dM^2 +\frac{1}{T I_M}dJ^2,
\end{eqnarray}
where $I_M$ is defined as 
\begin{equation}
I_M \equiv \beta \left(\frac{\partial J}{\partial \mu}\right)_M
 =\left [\frac{1}{2r_+^2}+\frac{J^2}{8Mr_+^4\triangle}
+\frac{J^2}{2M^2l^2r_+^2\triangle ^2}\right ]^{-1}.
\end{equation}
Through a straightforward calculation, we can easily verify
\begin{equation}
ds^2_{\rm R}=\beta ds^2_{\rm W},
\end{equation}
the conformal relation between the Ruppeiner and Weinhold metrics
\cite{Fer}. Because $C_J$ and $I_M$ are always positive, the line 
element $ds^2_{\rm R}$ is positive definite, so does $ds^2_{\rm W}$.
This is quite different from the case of the dilaton black holes in
N=2 supergravity theory \cite{Fer} (in the paper some speculations 
on the relation between the Weinhold geometry and the moduli space
geometry were given).

\subsection{Extremal Black Hole as the Critical Point}
In this section we see that the extremal black hole corresponds
to the critical point. One characteristic feature of the critical
point is the divergence of the correlation length. As is said in
the previous section, the thermodynamic curvature is proportional
to the correlation volume. Therefore the divergence of the 
thermodynamic curvature at some point suggests that it is the 
critical point.
Actually in the next section,
we will see for our case this divergence occurs at the extreme black 
hole. However, in order to confirm exactly that the extreme black 
hole is the critical point, we have to show that at least one of the 
second derivatives of the thermodynamic potential (appropriately 
chosen for the system at hand) diverges in the extreme limit, i.e., 
as $\triangle \rightarrow 0$. 
In fact, the critical point can be defined in several ways, 
here we follows the definition in \cite{Yeo}: the point where some 
of the second derivatives of the thermodynamic potential diverge.

As for the thermodynamic potential, we choose the Helmoltz free
energy $f=M-TS$. 
The Ruppeiner metric can be rewritten in the new coordinates $(T, J)$. 
\begin{eqnarray}
\label{rm4}
ds^2_{\rm R}&=&\frac{1}{T}(-\frac{\partial^2f}{\partial T^2}
+\frac{\partial^2f}{\partial J^2})\nonumber\\
&=&\frac{C_J}{T^2}dT^2+\frac{1}{TI_T} dJ^2,
\end{eqnarray}
where $C_J$ is given in (\ref{capacity}) and $I_T$ is
\begin{eqnarray}
I_T &\equiv & \left(\frac{\partial J}{\partial \Omega_H}\right)_T 
        \nonumber \\
	&=&\left [\frac{1}{2r_+^2}+\frac{J^2}{4Mr_+^4\triangle}+\left( 
	\frac{Jl^2(1+\triangle)}{4r_+^4}+
\frac{J^3}{4M^2r_+^4\triangle}\right)
	{\cal C}\right]^{-1}, 
\end{eqnarray}
with 
$$ {\cal C}=\left[\frac{J}{4 r_+^2}-\frac{J}
         { Ml^2\triangle}\right]\left[\triangle
			  +\frac{J^2}{M^2l^2 \triangle}-
			  \frac{Ml^2\triangle (1+\triangle)}{4 r_+^2}
			  -\frac{J^2}{4 Mr_+^2}\right]^{-1}. $$
We can easily see that $I_T$ vanishes in the extreme limit, in other
words, the second derivative of the Helmoltz free energy with respect 
to the angular momentum $J$ vanishes. This tells us that the extremal
black hole is the critical point and its Hawking temperature $T_H=0$
is the corresponding critical temperature. This zero 
temperature makes it hard to define the critical point in the usual
way as the point where some of the fluctuations diverge. This is why
we adopt the defintion given in \cite{Yeo}. In fact, as is mentioned 
before, the components of the Ruppeiner metric
have the meaning of the second moments for the fluctuation. (This 
interpretation is not valid at the critical point unless we take the
thermodynamic limit). The metric (\ref{rm4}) tells us in the 
thermodynamic limit that the following second moments vanish at the 
critical point. 
\begin{equation}
\label{sec3}
\langle \delta T \delta T\rangle_J =T^2/C_J, \ \
\langle \delta J \delta J\rangle_T =TI_T,
\end{equation}
where the subscripts denote the variables which 
are kept fixed. These are quite different results from the ordinary 
statistical system with non-vanishing critical temperature. There,
the fulctuaation diverges for the extensive quantities while it 
vanishes for the intensive quantities. 

\section{Thermodynamic Curvature}
In the different coordinates, however, the scalar curvature 
invariants should keep unchanged for the  same geometry. For a
two-dimensional Riemannian geometry, the most important curvature 
invariant is the scalar curvature $\Re$. For this purpose,  it is 
convenient to calculate it in the coordinates $(M, J)$.
In the following conventions, 
\begin{equation}
\Re^{\lambda}_{\mu\nu\sigma}=\Gamma ^{\lambda}_{\mu\nu, \sigma}-
\Gamma ^{\lambda}_{\mu\sigma, \nu}+\Gamma ^{\lambda}_{\sigma\eta}
\Gamma ^{\eta}_{\mu\nu} -\Gamma ^{\lambda}_{\nu\eta}
\Gamma ^{\eta}_{\mu\sigma},
\end{equation}
and 
\begin{equation}
\Re_{\mu\nu}=\Re^{\lambda}_{\mu\lambda\nu}, \ \ \Re=g^{\mu\nu}\Re_{\mu\nu},
\end{equation}
the  Riemannian scalar curvature of the Ruppeiner metric (\ref{rm1})
is
\begin{equation}
\label{curv}
\Re=\frac{1}{\sqrt{g}}\left [\frac{\partial}{\partial M}
\left (\frac{1}{\sqrt{g}}\frac{\partial g_{22}}
{\partial M}\right ) +\frac{\partial}{\partial J}
\left (\frac{1}{\sqrt{g}}\frac{\partial g_{11}}
{\partial J}\right ) \right ],
\end{equation}
where
\begin{equation}
g_{11}=1/(T^2 C_J), \ \, g_{22}=1/(T I_M).
\end{equation}
Further expansion of  expression for  the thermodynamic 
curvature (\ref{curv}) is somewhat complicated. Here we do not 
present it explicitly. Instead we demonstrate the behavior of the 
scalar curvature numerically.  
\begin{figure}[h]
\psfig{file=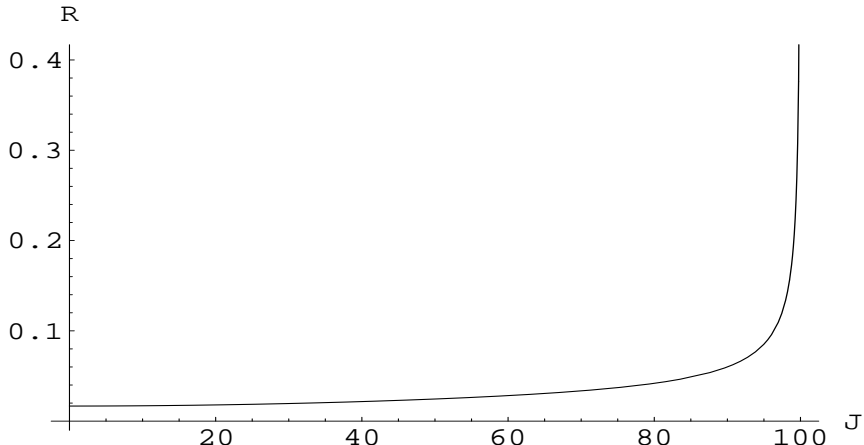,height=60mm,width=120mm,angle=0}
\caption{Thermodynamic curvature $\Re$ of the BTZ black hole 
thermodynamics 
versus the angular momentum $J$ with $M=1$ and $l=100$}
\end{figure}   
Fig. 1 shows the behaviour
of the scalar curvature with respect to the angular momentum with 
the mass and the cosmological constant set to be  
$M=1$ and $l=100$. Since the radius $l$ of the anti-de Stter space 
is concerned with the cosmological
constant, it is reasonable to set it lager than any other scale 
that appear in the theory. The thermodynamic 
curvature becomes small as $J \rightarrow 0$ but not vanishes. 
It means that even for the spinless black hole, the statistical
interaction is nontrivial in the corresponding
statistical model. It should be noted that the situation becomes
different if we initially work with the spinless black hole (setting
$J=0$). In that case, its thermodynamic geometry will be one dimensional
and the scalar curvature will vanish. Therefore the above nonvanishing 
scalar curvature is a result coming from another dimension specified
by $J$. When the extremal 
limit is approached, however, the curvature diverges strongly. 
More specifically, the inspection on the Eq.~(\ref{curv}) reveals 
that the curvature diverges near the extremal limit like 
\begin{equation}
\label{curv1}
\Re \sim \triangle ^{-1}.
\end{equation}

With the notion of the relation (\ref{scale}), we expect $\triangle$
to be related to the correlation length $\xi$.
Lacking of a satisfactory quantum theory of gravity, 
unfortunately, we do not have 
knowledge about the exact correlation function for the gravitational
interaction.
However, in Refs.\cite{Tra,Cai2,Cai3} it was 
already argued that the inverse surface gravity $\kappa=2\pi T$ of black 
holes may play the role of the correlation length
\begin{equation}
\xi=\frac{1}{2\kappa}.
\end{equation}
Since $\kappa\sim\triangle$ near the extremal point, 
combining (\ref{curv1}) and (\ref{scale}), we obtain the effective
spatial dimension of the near-extremal BTZ black holes;
\begin{equation}
\bar{d}=1.
\end{equation}
In fact, \cite{Cai3} argued from the scaling laws, which hold near 
the critical point, that the
effective spatial dimension of BTZ black holes is one.
Here we have further verified this result via the Riemannian 
curvature of the BTZ black hole thermodynamics. 

The divergence of the thermodynamic curvature, i.e., the 
correlation volume suggests that the extremal BTZ black hole
is the critical point. This can be expected from several other
aspects. For example, the extremal black hole corresponds to the 
BPS state in the supersymmetric extension. This means it preserves
some of the full supersymmetries of the supergravity, which is
not the case for the non-extremal black hole. Another fact is that
the Hawking temperature vanishes for the extremal black hole,
therfore no Hawking radiation happens. Moreover, beyond 
the extremality, the black hole becomes nakid. These discerning
properties of extremal black hole support the view point that
the extremal black hole is the critical point.  
 
Recalling the behavior of the thermodynamic 
curvature of ideal Boson gas \cite{Rup}, we find that the
behavior of the thermodynamic curvature for  the BTZ black holes is
similar to that of the Boson gas, but with an opposite sign. 
This signature difference looks strange, on which we will dicuss in
the later section. Here it should also be mentioned that more
recently Ghosh \cite{Gho} has constructed a one-dimensional Boson 
gas model to provide the correct expression for entropies for extremal 
and near-extremal BTZ black holes. Therefore our result is consistent 
with the one-dimensional gas model of the near extremal BTZ black hole.

It is interesting to see what would be the effective dimension of the 
statistical system corresponding to the non-extremal BTZ black hole.
One can extend the above procedure to get the effective dimension 
for arbitrary value of $J$, even far from the extremal point. 
The basic tool for this is the relation (\ref{scale}), which is
valid even far from the critical point. That is to say, from
\begin{equation}
\Re=c\left(\frac{l\sqrt{1+\triangle}}{2\sqrt{2M}\triangle}
\right)^{\bar{d}},
\end{equation}
we get
\begin{equation}
\bar{d}=\ln\frac{\Re}{c}/\ln\frac{l\sqrt{1+\triangle}}
{2\sqrt{2M}\triangle}, 
\end{equation}
the coefficient $c$ is determined by the 
condition that $\bar{d}=1$ at the critical point, which is clear
from the Eq. (\ref{curv1}). 

\begin{figure}[h]
\psfig{file=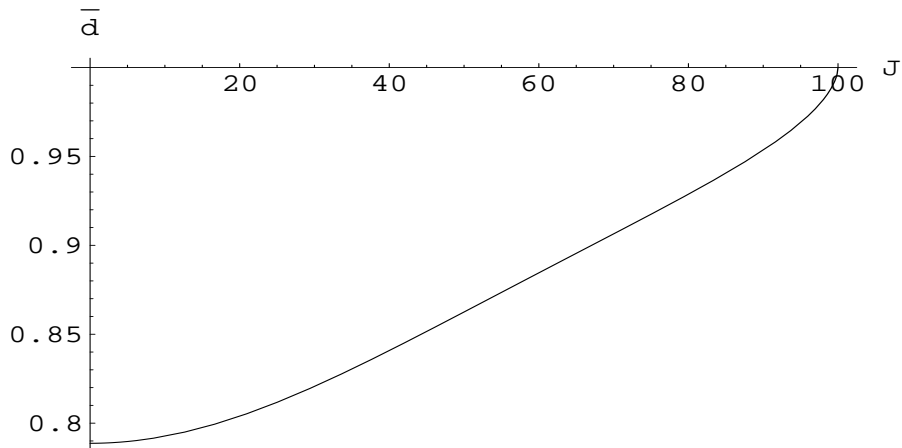,height=60mm,width=120mm,angle=0}
\caption{The effective dimension $\bar{d}$  
versus the angular momentum $J$ with $M=1$ and $l=100$. The 
coefficient $c$ is determined to be $3/(1250\pi)$.}
\end{figure}   
Fig. 2 shows the result, where we set $l=100$ and $M=1$ ($c$ is 
determined to be $3/(1250\pi)$). Remarkable
thing is that the effective dimension is not constant and 
becomes less than one as $J$ leaves the extremal point.
This variation of the effective dimension means that for each value
of the angular momentum $J$, the corresponding statistical model
should be constructed differently. This is in accord with the 
recent viewpoints on the black hole. In fact, for the different types
of black holes, their stringy counterparts are differently
suggested \cite{pol}. For the neutral black hole, the entropy 
can be matched with that of a single long string when its size 
becomes the string scale. As for the Ramond-Ramond charged black
hole, its stringy counterpart is the open string gas living on 
the D-brane. 

The fractal effective dimension looks absurd. The 
result becomes even worse when we choose the radius $l$ of the
anti-de Sitter space as the order of the black hole mass (the variation of 
thermodynamic curvature with respect to the angular momentum is insensitive
in shape to the value of $l$ as we see in Fig. 3).
In this case the effective dimension sharply decreases to the negative value
near the spinless black hole; see Fig. 4. 
\begin{figure}[h]
\psfig{file=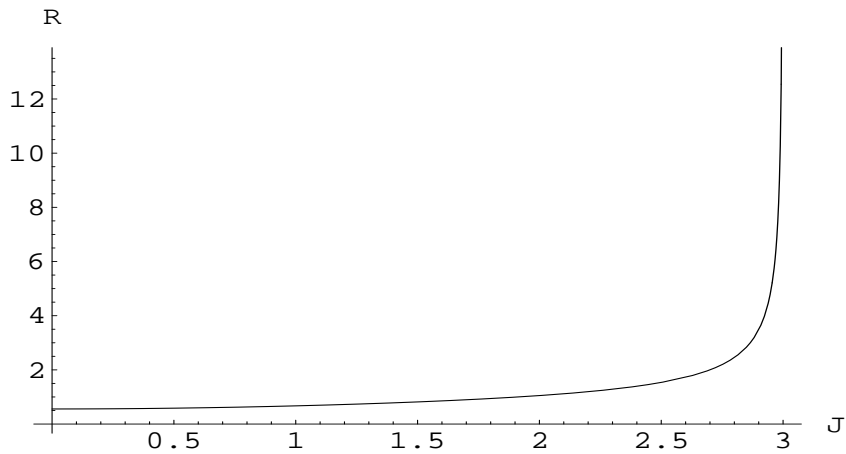,height=60mm,width=120mm,angle=0}
\caption{Thermodynamic curvature $\Re$   
versus the angular momentum $J$ with $M=1$ and $l=3$}
\end{figure}   
\begin{figure}[h]
\psfig{file=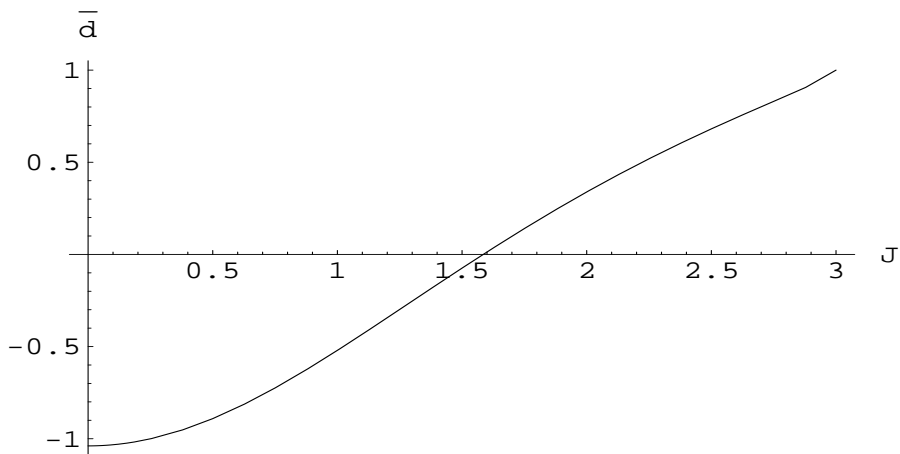,height=60mm,width=120mm,angle=0}
\caption{The effective dimension $\bar{d}$  
versus the angular momentum $J$ with $M=1$ and $l=3$. $c$ is determined to
be $8/(3\pi)$.}
\end{figure}   

Although it is very hard
to find such an example in the ordinary thermodynamic system, we can find one
in the string gas (bootstrap) model. 
This model has been dsicussed extensively concerning about the 
hadron behaviour at high density and high temperature. For example, 
the mean energy of some string gas model
behaves like $\langle E\rangle\sim (T_H-T)^{-1/2}$ near the 
Hagedorn temperature $T_H$ \cite{Fra}. If one try to interpret this 
result in terms of ordinary particle gas, which behaves like 
$\langle E\rangle\sim T^{\bar{d}+1}$ in $\bar{d}$ spatial 
dimension, he will get 
the negative effective dimension $\bar{d}=-3/2$. Of course, the 
string gas does not live in the negative spatial dimension.
In this case, the negative effective dimension comes from the 
extrapolation of the interpretation of the particle system to 
the string gas. 
Therefore we have at least one possibity: The corresponding statistical 
model is not an ordinary sytem composed of particle-like gas. 
We think that some extended objects constitutes the system. 

\section{Discussions}
In the thermodynamics of BTZ black holes we have introduced the 
two-dimensional Ruppeiner metric in various coordinates. This metric
is always positive definite. For general black holes, due to the fact
that the heat capacity of some black holes may be negative, the
Ruppeiner metric of black holes is not always positive definite. 
For example, for the Kerr black hole,
the counterpart of the BTZ black hole in four dimensions, its heat
capacity is positive for large angular momentum, and negative for small
angular momentum, that is, the heat capacity changes its sign at some
points in the parameter space. Therefore, the Ruppeiner metric for the
Kerr black holes is not always positive definite. But near the critical
points, the positive definiteness of the Ruppeiner metric is
guaranteed. 

Using the Ruppeiner metric we investigated 
the critical point for the BTZ black holes 
in the different thermodynamic coordinates. The divergence of 
the Riemannian curvature 
of the Ruppeiner metric suggests the existence of a critical 
point at the extremal limit of  BTZ black holes. This is verified
via the second derivative of the Helmoltz free energy
with respect to the angular momentum.

Now we come back to the subtle point we mentioned above: 
the signature flip of the curvature. As is said before, Ghosh
recently found \cite{Gho} the relation between the near extremal 
BTZ black hole 
and $(1+1)$-dimensional conformal field theory via the ideal bosonic
gas model in one dimension. This strongly imples that the system
is composed of bosons. However, the ideal quantum gas gives 
negative curvature for boson. This point is quite unclear to us but
one definite thing is that our reslut reveals sharp divergence
near the critical point, which is one characteristic of boson 
showing the condensation. 

One way out is that the 
criterior based on the signature of the scalar curvature might
not be valid for the sytem of the extended object. In fact,
we can find one example in the Takahashi gas. This is the one
dimensional system composed of rigid rods. In this model, the 
scalar curvature shows sharp peak at a point considered as
the pseudo-phase transition point in the sense that it is still
finite (See fig. 11 of the \cite{Rup}, with the notion that there
is drawn the Gaussian curvature which has the opposite signature
to that of Riemannian curvature). One can easily see that it shows
negative Gaussian curvature, i.e., positive scalar curvature 
in the region of liquid phase (larger density region). 

Another interesting thing is that the curve in this liquid phase region 
shows very similar shape to the our result.
We think this is not a coincidence. If we grow the size of the rigid
rod, the density becomes large and system will be in the 
liquid phase from some point. Further increase of the rod length
terminates with zero scalar curvature at the point where the whole 
system is composed of one long rigid rod. Of course, this cannot
be the exact statistical model corresponding to the BTZ black hole
because the rigidity is too strong to give a nonvanishing
curvature at the spinless black hole limit. Nevertherless the 
qualitative feature can be the same and one can 
view our result in the same way; Near the extremal point, the system 
is composed of very very short string gas. With the decrease of 
the angular momentum $J$, the size of the string gas becomes longer. 
Near the spinless black hole point, it becomes highly 
oscillatory single long string. In fact, this viewpoint has been
discussed in many papers in different contexts but our results
provide a strong support for the viewpoint.

\section*{acknowledgements}
This work was supported in part by the Center for Theoretical Physics
of Seoul National University. JHC thanks Dr. J. Lee for 
helpful discussion on the statistical models and he is also supported 
by the Korea Science and Engineering Foundation.

\end{document}